\documentclass[%
 reprint,
superscriptaddress,
 amsmath,amssymb,
 aps,
prapplied,
doi=true,
floatfix,
]{revtex4-1}

\usepackage[colorlinks=true,linkcolor=blue,urlcolor=blue,citecolor=blue]{hyperref}
\usepackage{graphicx}
\usepackage{dcolumn}
\usepackage{bm,upgreek}
\usepackage{overpic}
\usepackage[caption=false]{subfig}
\usepackage{comment}
\usepackage{xfrac}
\usepackage{xcolor}
\usepackage{dblfloatfix}
\usepackage{mathtools}
\usepackage{CJK}
\usepackage{bm}
\usepackage{multirow}
\usepackage{gensymb}
\usepackage{pifont}
\usepackage[mathscr]{euscript}

\usepackage[export]{adjustbox}

\usepackage{amssymb}
\usepackage{relsize}
\usepackage[normalem]{ulem}

\definecolor{dandelion}{rgb}{0.94, 0.88, 0.19}
\definecolor{thepurple}{rgb}{0.65, 0.24, 0.59}
\definecolor{theorange}{rgb}{0.95, 0.40, 0.13}
\definecolor{thegreen}{rgb}{0.05, 0.50, 0.25}
\definecolor{mathematicagreen}{rgb}{0.0, 0.80, 0.0}
\definecolor{igoraquamarine}{rgb}{0.0, 0.6, 0.6}
\definecolor{igorgreen}{rgb}{0.0, 0.6, 0.0}
\definecolor{springgreen}{rgb}{0.0, 0.8, 0.0}
\definecolor{CaptAlsid}{rgb}{1.0,0.4,0}
\definecolor{skyblue}{rgb}{0.0, 0.6, 1.0}
\definecolor{black}{rgb}{0.0 0.0 0.0}

\begin{document}

\preprint{APS/123-QED}

\begin{CJK*}{UTF8}{}
\title{A Solid-State Microwave Magnetometer with Picotesla-Level Sensitivity}

\author{Scott T. Alsid}
\affiliation{Lincoln Laboratory, Massachusetts Institute of Technology, Lexington, Massachusetts 02421, USA}
\affiliation{Research Laboratory of Electronics and Department of Nuclear Science and Engineering, Massachusetts Institute of Technology, Cambridge, Massachusetts 02139, USA}
\author{Jennifer M. Schloss}
\email[Corresponding author.\\]{jennifer.schloss@ll.mit.edu}
\affiliation{Lincoln Laboratory, Massachusetts Institute of Technology, Lexington, Massachusetts 02421, USA}
\author{Matthew H. Steinecker}
\affiliation{Lincoln Laboratory, Massachusetts Institute of Technology, Lexington, Massachusetts 02421, USA}
\author{John F. Barry}
\affiliation{Lincoln Laboratory, Massachusetts Institute of Technology, Lexington, Massachusetts 02421, USA}
\author{Andrew C. Maccabe}
\affiliation{Lincoln Laboratory, Massachusetts Institute of Technology, Lexington, Massachusetts 02421, USA}
\author{Guoqing Wang \CJKfamily{gbsn}(王国庆)}
\affiliation{Research Laboratory of Electronics and Department of Nuclear Science and Engineering, Massachusetts Institute of Technology, Cambridge, Massachusetts 02139, USA}
\author{Paola Cappellaro}
\affiliation{Research Laboratory of Electronics and Department of Nuclear Science and Engineering, Massachusetts Institute of Technology, Cambridge, Massachusetts 02139, USA}
\affiliation{Department of Physics, Massachusetts Institute of Technology, Cambridge, Massachusetts 02139, USA}
\author{Danielle A. Braje}
\affiliation{Lincoln Laboratory, Massachusetts Institute of Technology, Lexington, Massachusetts 02421, USA}
\maketitle
\end{CJK*} 

\date{\today}

\begin{abstract}
Quantum sensing of low-frequency magnetic fields using nitrogen-vacancy (NV) center ensembles has been demonstrated in multiple experiments with sensitivities as low as $\sim$1~pT/$\sqrt{\text{Hz}}$.
To date, however, demonstrations of high-frequency magnetometry in the GHz regime with NV diamond are orders of magnitude less sensitive, above the~nT/$\sqrt{\text{Hz}}$ level. Here we adapt for microwave frequencies techniques that have enabled high-performance, low-frequency quantum sensors. Using a custom-grown NV-enriched diamond combined with a noise cancellation scheme designed for high-frequency sensing, we demonstrate a Rabi-sequence-based magnetometer able to detect microwave fields near 2.87 GHz with a record sensitivity of 3.4~pT/$\sqrt{\textrm{Hz}}$. We demonstrate both amplitude and phase sensing and project tunability over a~300 MHz frequency range. This result increases the viability of NV ensembles to serve as microwave circuitry imagers and near-field probes of antennas. 
\end{abstract}

\pacs{Valid PACS appear here}
\maketitle

\section{\label{sec:intro}Introduction}
\begin{figure}
  \centering
        \includegraphics[width=0.48\textwidth]{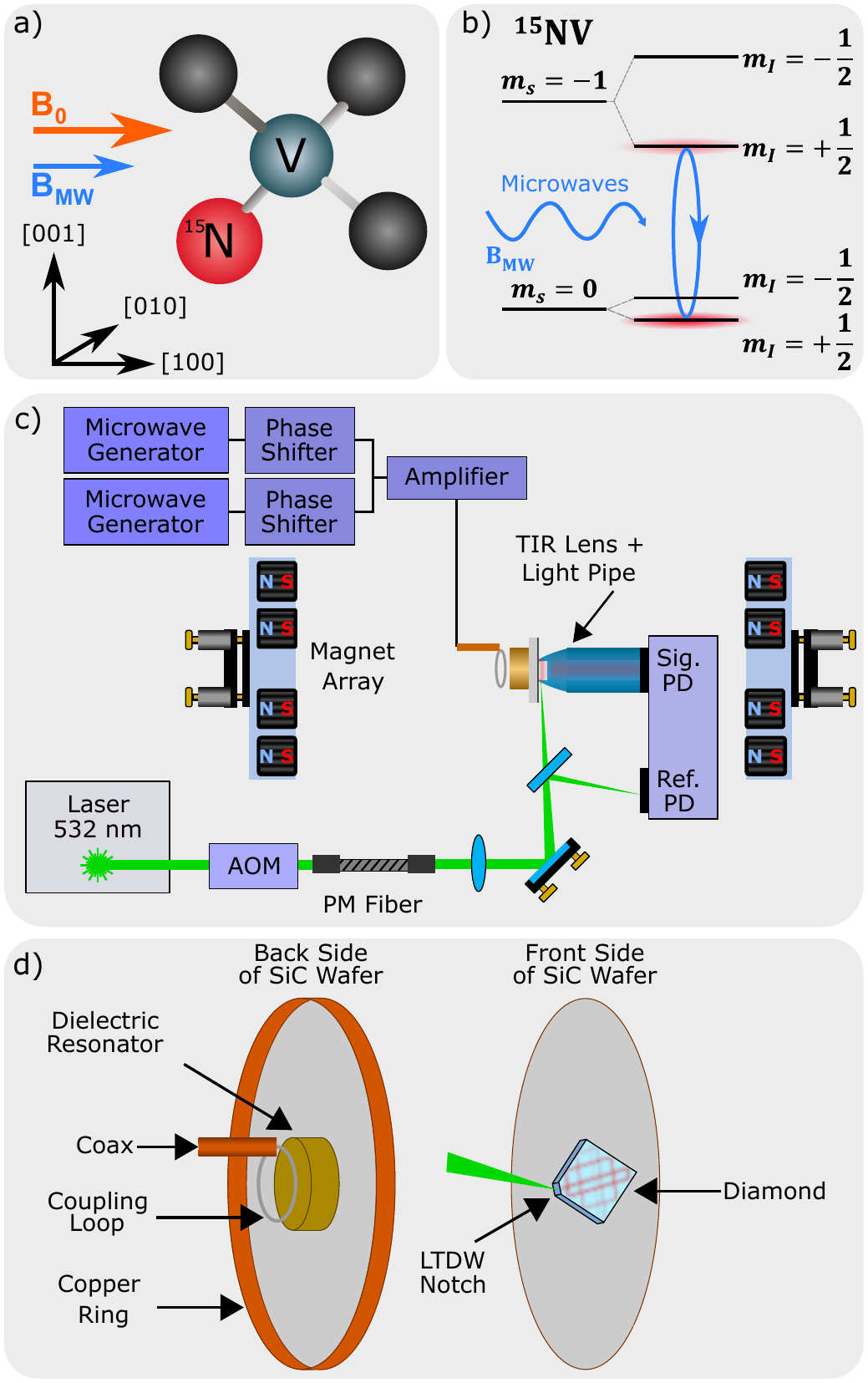}
            \caption{\small \textbf{Experimental setup for Rabi magnetometry using NV centers in diamond.} (a) The NV center. The static field $B_0$ and microwave (MW) field $B_{\textrm{MW}}$ are aligned along the diamond's $\langle 100 \rangle$ direction, projecting equally onto (and transverse to) all four crystollographic axes, resulting in the same resonance frequency and equal MW driving strength for all four NV orientations. (b) Energy level diagram for $^{15}$NV showing a resonant MW field driving Rabi oscillations between the \mbox{$|m_s \!=\! 0, m_I \!=\! +\frac{1}{2}\rangle$} and \mbox{$|m_s \!=\! -1, m_I \!=\! +\frac{1}{2}\rangle$} states. (c) Schematic of the experimental apparatus. (d) Detailed schematic of the silicon-carbide diamond mount and MW delivery system. }
        \label{fig:ACExperimentalSetup}
\end{figure}

Solid-state spin systems are increasingly favored for quantum sensing~\cite{Degen2017}. With resonance shifts tied to physical constants~\cite{Budker2007}, these robust systems can be tailored to applications covering a wide range of physical conditions while offering sensitivity and stability~\cite{Fu2020}. 
Nitrogen-vacancy (NV) centers in diamond [Fig.~\ref{fig:ACExperimentalSetup}(a)] constitute a particularly promising solid-state platform~\cite{Doherty2013}, with spin lifetimes exceeding milliseconds~\cite{Balasubramanian2009}, high spatial-resolution optical readout~\cite{LevineTurner2019}, intrinsic vector sensing capabilities~\cite{Schloss2018Vector}, and compatibility with ambient or extreme temperatures~\cite{Toyli2012} and pressures~\cite{Doherty2014Pressure}. These advantages have allowed demonstrations of low- and intermediate-frequency sensing applications such as non-invasive detection of neuron action potentials~\cite{Barry2016}, high-resolution NMR~\cite{Glenn2018}, single protein detection \cite{Lovchinsky2016}, elucidation of condensed matter phenomena~\cite{Casola2018}, mapping of remnant magnetization in geological samples~\cite{Glenn2017}, and magnetic navigation~\cite{Fleig2018MagNav,Frontera2019MagNav}. Performance in many of these applications is limited by the device's magnetic field sensitivity~\cite{Barry2020}.  The best reported sensitivities of NV-diamond magnetometers approach or surpass 1~pT/$\sqrt{\textrm{Hz}}$ for near-DC fields up to several kilohertz~\cite{Fescenko2020,Eisenach2021CR,barry2022sensitive}, and tens of pT/$\sqrt{\textrm{Hz}}$ for AC fields from roughly 100~kHz to a few megahertz~\cite{LeSage2012,Wolf2015,Glenn2018}.\par

Although NV-diamond magnetometry efforts have traditionally focused on sensing sub-MHz fields, operation at microwave (MW) frequencies could expand the application space to include imaging of MW circuitry components~\cite{Anderson2018Imaging}, antenna characterization~\cite{Jomaa2017RFIDAntenna}, and wireless communications~\cite{Meyer2018}.  
To date, however, the best demonstrated NV-diamond-based sensitivities at MW frequencies are orders of magnitude worse than those achieved at low frequency, with reported values in the 10s of nT/$\sqrt{\textrm{Hz}}$ range or above~\cite{Shao2016,Horsley2018}. This poor sensitivity results from the low fluorescence signal collection efficiency, limited NV-interrogation volume, and MW inhomogeneities present in typical high-frequency NV sensing experiments. \par

Here we demonstrate a high-frequency NV-diamond magnetometer tuned to detect amplitude modulation of a 2.863~GHz MW field. The sensor employs a Rabi-based measurement scheme in a diamond exhibiting long spin dephasing times and low strain. Sensitivity to non-magnetic noise is mitigated by inverting and subtracting the magnetometry signal from consecutive measurements using an additional MW pulse. Combining this scheme with recently demonstrated improvements in optical collection efficiency, laser noise cancellation, and uniform MW delivery~\cite{barry2022sensitive}, the resultant device exhibits a sensitivity of 3.4~pT/$\sqrt{\textrm{Hz}}$ for AM modulation of a MW carrier over the 100~Hz to 3.3~kHz band.




\section{Rabi Magnetometry}\label{sec:RabiNutations}
When an ideal two-level system is driven by a near-resonant magnetic field~\cite{Rabi1937SpaceQuantization,Jeleko2004Coherent}, the population oscillates between the two states $|0\rangle$ and $|1\rangle$ such that the population $P_{|1\rangle}$ is given by the Rabi formula,  \begin{equation}\label{eq:RabiForm}
P_{|1\rangle}(\tau) = \frac{\Omega^2}{\Omega^2+\delta^2}\sin^2\!\left(\frac{\sqrt{\Omega^2+\delta^2} \tau}{2}\right),
\end{equation}
where for the frequency $\omega$ of the driving field, $\delta = \omega-\omega_0$ is the detuning from the resonance frequency $\omega_0$, and $\tau$ is the duration of the driving field. For a coherent MW field $B_{\textrm{MW}}$, we have the Rabi frequency $\Omega\sim \gamma B_{\textrm{MW}}$, where $\gamma$ is the gyromagnetic ratio ($\gamma = 2\pi \times 2.8$~MHz/G for an electron spin).\par

Rabi magnetometry exploits these population oscillations: small changes in the near-resonant magnetic field amplitude cause the observed oscillation frequency to vary. This variation can be detected by sweeping $\tau$ and recording a set of Rabi fringes to determine changes in $\Omega$.  Alternatively, $\tau$ may be fixed and the signal monitored to detect sufficiently small changes in $P_{|1\rangle}$. While Eq.~\eqref{eq:RabiForm} describes the ideal behavior, Rabi oscillations in a real ensemble decay at a characteristic time scale $T_{2\rho}$, which is the driven-evolution analogue of the free-evolution dephasing time $T_2^*$~\cite{Geva1995}. The observed value of $T_{2\rho}$ depends on various dephasing mechanisms in the diamond lattice and spatial inhomogeneities in the drive field intensity~\cite{Cai2012RobustDD}. \par






The Rabi magnetometry scheme employed here consists of a resonant MW field applied to an ensemble of NV centers in diamond [see Fig.~\ref{fig:ACExperimentalSetup}(a,b)] for fixed duration $\tau$, followed by spin-state-dependent optical readout~\cite{Jeleko2004Coherent}. Monitoring the output signal for each experimental repetition allows amplitude changes in the MW field over time to be measured. The shot-noise-limited MW magnetic field sensitivity $\eta_\text{shot}$ of a Rabi magnetometer using $N$ NV centers is given by~\cite{Alsid2021RabiSupplement} 
\begin{equation}
\label{eq:sensitivity}
\eta_\text{shot} \approx \frac{\sqrt{2}}{\gamma}\frac{1}{F_{\perp}} \frac{1}{\sqrt{Nn_{\textrm{avg}}}}\frac{1}{Ce^{-(\tau/T_{2\rho})^p}}\frac{\sqrt{\tau+t_O}}{\tau}.
\end{equation}
Here $C$ is the fluorescence contrast between spin states, $n_{\textrm{avg}}$ is the average number of photons collected per NV per measurement,  $F_{\perp}$ is the projection of the applied MW magnetic field on the plane perpendicular to the NV axis, and $p$ is a stretched exponential parameter. The overhead time $t_O$ is the duration of any additional steps in the sequence beyond the MW pulse, including optical initialization and readout.  Assuming optimal choice of $\tau$, the performance of the device depends on the measurement contrast $C$, the value of $T_{2\rho}$, the number of NVs addressed, and the value of $n_{\text{avg}}$~\cite{Barry2020}. 

\section{Experimental Design}
We now highlight experimental design choices to achieve high-performance magnetometry. The experimental setup is shown schematically in Fig.~\ref{fig:ACExperimentalSetup}(c). This device is modified from the low-frequency magnetometer in Ref.~\cite{barry2022sensitive}; for additional experimental setup details pertinent to the present demonstration, see the Supplemental Material~\cite{Alsid2021RabiSupplement}. The sensor uses a $~3\;\textrm{mm} \times 3\;\textrm{mm}\times 0.62\;\textrm{mm}$ single-crystal diamond with $\langle 100 \rangle$ sides and an approximately 70-$\mu$m-thick $^{15}$N-doped layer, referred to as the NV layer. The diamond is adhered to a SiC wafer, which acts as a heatsink.  Laser light at 532~nm excites NVs throughout the NV layer, and the resulting NV fluorescence is collected by a photodiode. A 2.23~gauss bias field splits out the NV electron spin resonances, allowing the \mbox{$|m_s \!=\! 0, m_I \!=\! +\frac{1}{2}\rangle \leftrightarrow |m_s \!=\! \pm 1, m_I \!=\! +\frac{1}{2}\rangle$} transitions to be resolved spectroscopically.\par

The diamond is custom-grown, and the 70-$\mu$m-thick NV layer exhibits a measured 99.998$\%$~$^{12}$C purity. The diamond is grown using the $^{15}$N isotope.  The $^{15}$NV centers exhibit only two hyperfine features per spin resonance and a larger splitting compared to the $^{14}$NV center, which has three hyperfine features. 
A single MW tone can thus drive a larger fraction of the $^{15}$NV population and is less likely to induce off-resonant driving of other hyperfine features. Moreover, the diamond exhibits strain variation less than $2\pi \times 10$~kHz. This yields a $T_2^* \approx 9\;\mu$s dephasing time~\cite{barry2022sensitive}, which is longer than the typical $T_2^*\lesssim 1\;\mu$s encountered in NV ensemble magnetometry~\cite{Barry2020}. The resulting resonance linewidth is $\Gamma = 2/T_2^* \approx 2\pi \times 35$~kHz. The narrow ensemble linewidth results in an extended $T_{2\rho}$ by (a) reducing the spread of individual NV detunings from the MW drive frequency and (b) enabling the NV ensemble to be driven effectively at a lower Rabi frequency, reducing the spread of MW field strengths across the ensemble due to any MW inhomogeneity.

The device here employs the light trapping diamond waveguide technique \cite{Clevenson2015}, which excites NVs throughout the entire $~3\;\textrm{mm} \times 3\;\textrm{mm}\times 70\;\mu\textrm{m}$ NV-doped volume [Fig.~\ref{fig:ACExperimentalSetup}(d)]. Fluorescence collection efficiency is increased to approximately 75-95$\%$ by surrounding the diamond with a total-internal-reflection lens~\cite{barry2022sensitive}. In addition, laser light is sampled immediately prior to entering the diamond and is directed to a reference photodiode for common-mode noise cancellation~\cite{Alsid2021RabiSupplement}.\par

Both the bias magnetic field $B_0$ and MW field $B_{\textrm{MW}}$ are oriented normal to the diamond's $\{100\}$ face. $B_0$ projects equally onto all four NV symmetry axes [see Fig.~\ref{fig:ACExperimentalSetup}(a)], splitting the \mbox{$|m_s \!=\! 0\rangle \leftrightarrow |m_s \!=\! \pm1\rangle$} transitions equally. Meanwhile, $B_{\textrm{MW}}$ projects equally onto the planes transverse to the four NV axes. As a result, all NV classes may be addressed with a single MW tone at the same Rabi frequency. 
Because the laser excites NVs throughout the NV layer, it is necessary to minimize static and MW field gradients over the entire NV-doped volume to avoid degrading $T_{2\rho}$~\cite{deSousa2009}. The bias magnet arrangement is designed to minimize static field gradients, while inhomogeneities in $B_{\textrm{MW}}$ are suppressed by coupling MWs from a shorted coaxial cable to a dielectric resonator to achieve a greater than 90$\%$ uniformity over the $^{15}$NV layer (see the Supplemental Material~\cite{Alsid2021RabiSupplement}).\par


\begin{figure}
  \centering
        \includegraphics[width=0.45\textwidth]{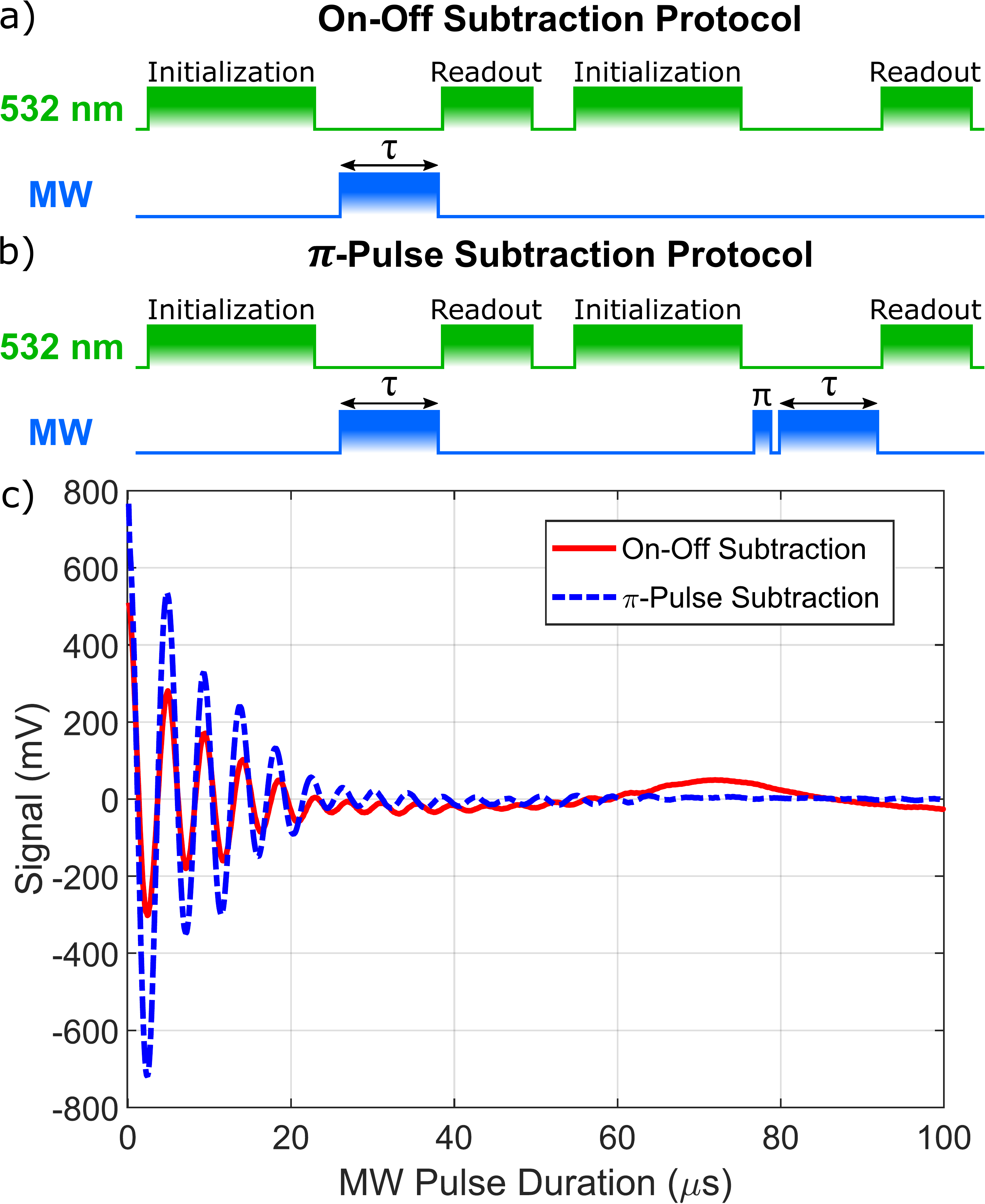}
            \caption{\small \textbf{Noise Cancellation.} (a) The on-off subtraction protocol consists of two subtracted sequences with a MW drive only in the first sequence. (b) The $\pi$-pulse subtraction protocol consists of two subtracted Rabi sequences, where the second sequence contains an additional MW $\pi$~pulse. While both protocols help eliminate low frequency noise; the $\pi$-pulse protocol maximizes the magnetometry measurement signal-to-noise ratio (SNR) and suppresses the $^{15}$N hyperfine modulation observed in the Rabi oscillations. (c) Rabi oscillations at 220~kHz for each subtraction protocol using a resonant drive field. The $\pi$-pulse protocol suppresses the hyperfine-mediated modulation, allowing for a straightforward extraction of $T_{2\rho}$ via a decaying sinusoid fit.}
        \label{fig:RabiandHyperfine}
\end{figure}

A single Rabi sequence begins with optical initialization of the NV ensemble, followed by application of a resonant MW pulse of duration $\tau$, and finally optical readout of the NV state.  This process can be repeated while varying $\tau$ to observe the Rabi oscillations and decay envelope. In some Rabi magnetometry implementations~\cite{Horsley2018}, each Rabi sequence is followed by a reference sequence with no MWs applied. The signal from the Rabi measurement is then normalized using the signal from the reference sequence, as shown in Fig.~\ref{fig:RabiandHyperfine}(a).  This ``on-off'' subtraction protocol suppresses low-frequency noise at the expense of reduced measurement bandwidth and signal, as half of all measurements produce zero contrast.  To improve SNR, we design and implement a subtraction scheme that consists of repeated Rabi sequences performed with and without a $\pi$-pulse applied before the Rabi drive pulse; the signals from successive sequences are then subtracted to produce the magnetometry signal, as shown in Fig.~\ref{fig:RabiandHyperfine}(b). This ``$\pi$ pulse'' subtraction scheme mitigates the signal reduction while still suppressing low-frequency noise.\par 

The MW field drives a transition to the hyperfine-resolved $|m_s \!=\! -1, m_I \!=\! +\frac{1}{2}\rangle$ state, but the near-degeneracy of the $|m_s \!=\! 0, m_I \!=\! -\frac{1}{2}\rangle$ and $|m_s \!=\! 0, m_I \!=\! +\frac{1}{2}\rangle$ hyperfine ground states in a small static field~\cite{Felton2009Hyperfine} complicates the interpretation of the measured Rabi oscillations~\cite{oon2022ramsey}. The Larmor precession between these near-degenerate states results in a modulation of the overall Rabi oscillation signal as seen in Fig.~\ref{fig:RabiandHyperfine}(c).  The $\pi$-pulse subtraction scheme suppresses this hyperfine-mediated modulation arising from the presence of the $^{15}$N nucleus in the $^{15}$NV center~\cite{gaebel2006room,ohno2012engineering}, which also allows a more robust determination of $T_{2\rho}$ (see the Supplemental Material~\cite{Alsid2021RabiSupplement}). 

\section{Magnetic Field Sensitivity}\label{sec:MagneticFieldSensitivity}
\begin{figure}
  \centering
        \includegraphics[width=0.45\textwidth]{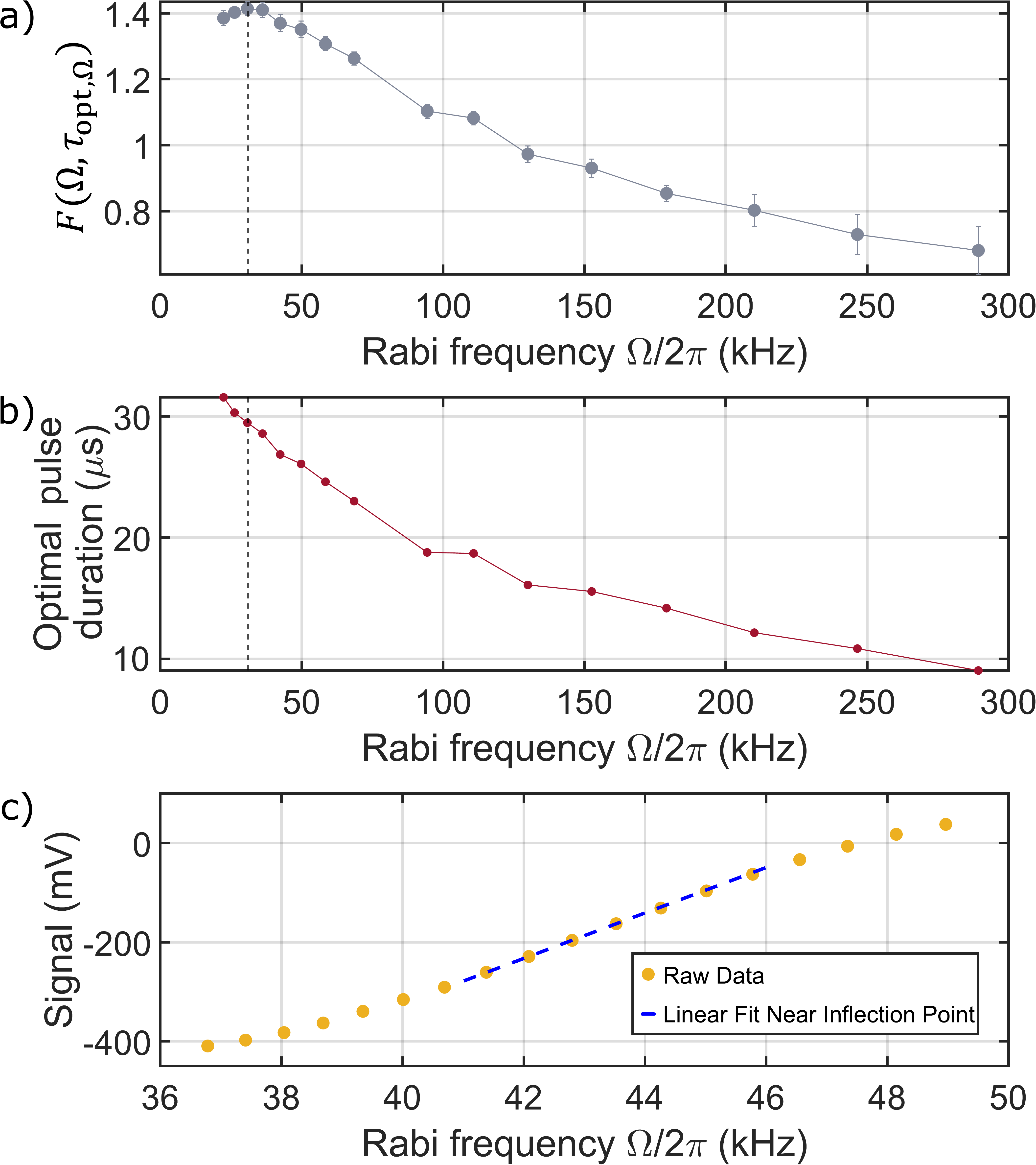}
            \caption{\small \textbf{Rabi Frequency Dependence of Key Device Parameters.} (a) Sensitivity figure of merit $F(\Omega,\tau_{\text{opt},\Omega})$ at different MW field strengths for a $100\;\mu$s Rabi pulse sequence, which produces the initial estimate of the optimal Rabi frequency $\Omega_{\textrm{opt}}$. (b) Optimal MW pulse duration $\tau_{\text{opt},\Omega}$ calculated to maximize the figure of merit $F(\Omega,\tau_{\text{opt},\Omega})$ for a given Rabi frequency $\Omega$. The value corresponding to the maximum value of $F(\Omega,\tau_{\text{opt},\Omega})$ sets the MW pulse length for the magnetometry sequence.  (c) Empirical determination of the optimum Rabi frequency for a fixed MW duration $\tau = 30\;\mu$s. The NV ensemble fluorescence signal is most sensitive to changes in the applied MW field at the point of steepest slope, $\Omega_{\textrm{opt}} = 2\pi \times 42.8$~kHz.} 
        \label{fig:T2rhoandRFscan}
\end{figure}



The Rabi frequency that optimizes the sensitivity in Eq.~\eqref{eq:sensitivity} is set by a balance of contrast $C$ and driven coherence time $T_{2\rho}$. Lowering $\Omega$ increases $T_{2\rho}$ by reducing the effects of MW field gradients and noise, but also degrades the contrast $C$, as a smaller fraction of the inhomogeneously-broadened ensemble is addressed. To quantify this balance, we derive a figure of merit $F(\Omega,\tau)$ from Eq.~\eqref{eq:sensitivity},
\begin{equation}\label{eq:RabiFOM}
F(\Omega,\tau) = C(\Omega) e^{-\tau/T_{2\rho}(\Omega)}\frac{\tau}{\sqrt{\tau + t_O}},
\end{equation}
where $\tau$ is the duration of a MW pulse with intensity $\Omega$; $t_O = 44.8\;\mu$s is the sequence overhead time due to the laser initialization and readout pulses, the $\pi$-pulse, and electronic dead time; and $C(\Omega)$ and $T_{2\rho}(\Omega)$ are the measured contrast and dephasing time, respectively, as functions of the Rabi frequency. Here maximizing $F(\Omega,\tau)$ optimizes sensitivity (minimizes the value of $\eta$).  

We measure a series of Rabi oscillation curves, sweeping $\tau$ from 0 to 100~$\mu$s in a fixed-length Rabi sequence for a range of values of $\Omega$. We fit each curve to extract $T_{2\rho}$ and $C$~(see Fig. S4 of the Supplementary Material~\cite{Alsid2021RabiSupplement}).  Using these measurements and Eq.~\eqref{eq:RabiFOM}, we determine the optimal MW pulse duration $\tau_{\text{opt},\Omega}$ that maximizes $F(\Omega, \tau)$ for a given $C(\Omega)$ and $T_{2\rho}(\Omega)$. Plotting the values of $F(\Omega,\tau_{\text{opt},\Omega})$ as a function of $\Omega$, we find that $\Omega = 2\pi \times 31$~kHz maximizes the figure of merit $F(\Omega,\tau_{\text{opt},\Omega})$ with $\tau_{\text{opt},\Omega} = 30$~$\mu$s, as shown in Fig.~\ref{fig:T2rhoandRFscan}(a,b). We thus generate a new Rabi sequence by fixing $\tau = 30$~$\mu$s, which eliminates excess overhead time between the initialization and readout pulses. To account for small changes in experimental conditions when operating this shorter sequence, we perform a final optimization, varying $\Omega$ to find the value that maximizes the slope of the fluorescence signal $S$, $\left|\partial S/\partial \Omega\right|$, as shown in Fig.~\ref{fig:T2rhoandRFscan}(c). This optimal Rabi frequency $\Omega_\text{opt}$ maximizes the response to a given magnetic field; we find the $\tau=30\;\mu$s sequence exhibits an optimal Rabi frequency of $\Omega_{\textrm{opt}} = 2\pi \times 42.8$~kHz.

The $\tau = 30\;\mu$s Rabi sequence is performed at a 13.23~kHz rate. With the $\pi$-pulse subtraction protocol, magnetometry data are collected at a 6.615~kHz rate. The measurement is run continuously to yield a 1-second-length fluorescence time series. Each time series is Fourier transformed with rectangular windowing to produce a single-sided amplitude spectral density (ASD) with 1-Hz-wide frequency bins up to 3.3075~kHz.\par

\begin{figure}
  \centering
        \includegraphics[width=0.45\textwidth]{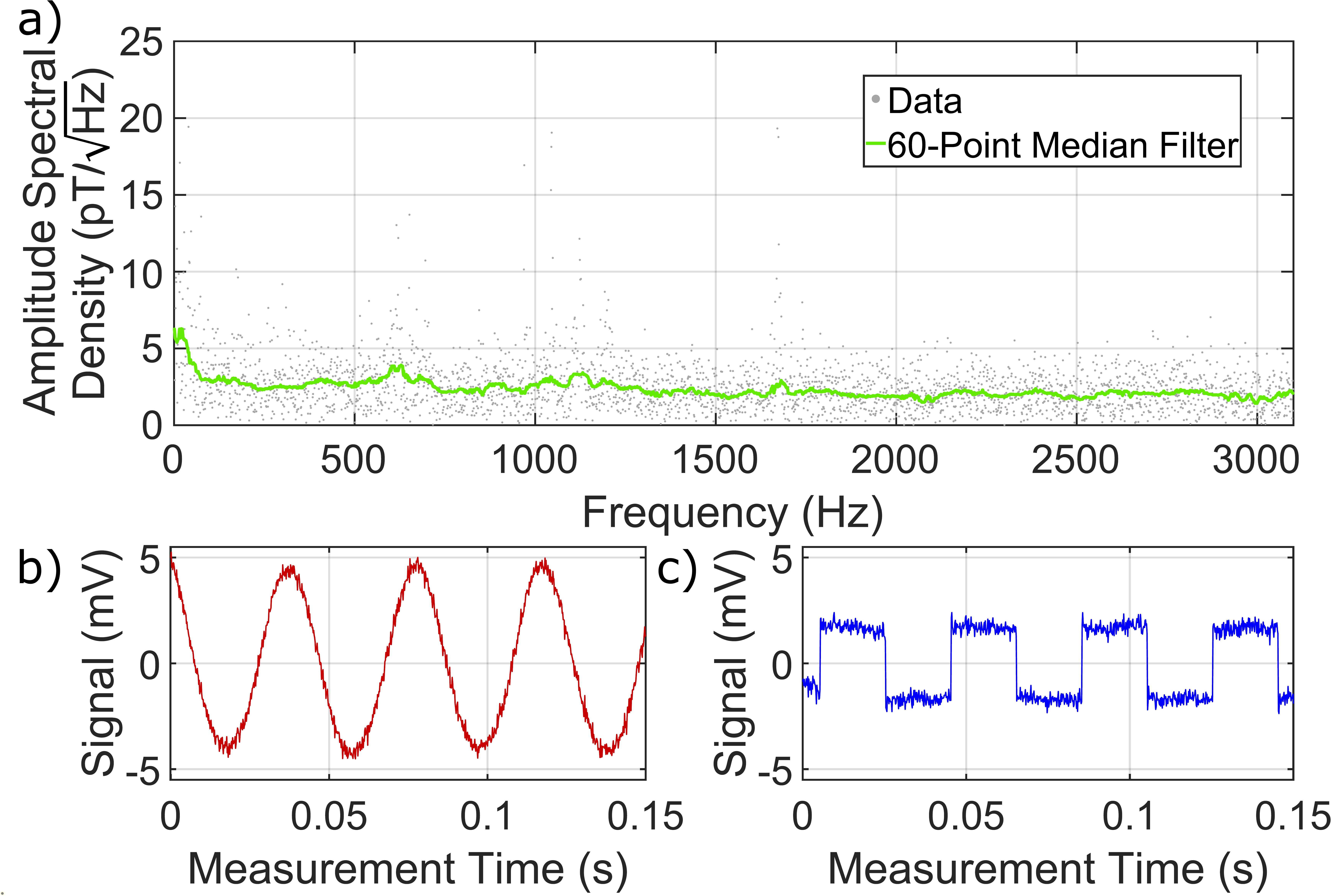}
            \caption{\small \textbf{Rabi magnetometry.} (a) Amplitude spectral density of a 1-second measurement with no signal field applied. The units are calibrated by the slope and the NV gyromagnetic ratio as discussed in the main text. (b) Measurement of a separate $\delta B_{\textrm{MW}} = 4.4$~nT, 25-Hz amplitude modulated magnetic field with 100$\%$ modulation depth. (c) Measurement of a separate $\delta B_{\textrm{MW}} = 2.2$~nT, 25-Hz phase modulated magnetic field with a $\pi/4$ phase deviation.}
        \label{fig:RabiMagSignalASD}
\end{figure}

We next determine the sensitivity $\eta$ 
for a modulation frequency band of interest.  Assuming white Gaussian noise in this band, the sensitivity can be expressed as
\begin{equation}
    \eta = \frac{\sigma_B}{\sqrt{2 f_{\textrm{ENBW}}}},
\end{equation}
where $\sigma_B$ is the standard deviation of the zero-signal magnetic-field time trace over the equivalent noise bandwidth $f_{\textrm{ENBW}}$~\cite{Schloss2018Vector}. We first calculate the standard deviation $\sigma_S$ of a time series of fluorescence measurements by applying a 100-Hz brick-wall high-pass filter (such that $f_{\text{ENBW}} = 3207$~Hz) on the amplitude spectral density and then finding the root-mean-square average. The standard deviation $\sigma_S$ is then divided by the slope $\left|\partial S/\partial \Omega\right|$ and $\gamma$ to express the standard deviation in magnetic field units, $\sigma_B$.  Taking into account the projection of the $\langle100\rangle$-directed MW field onto the plane perpendicular to each $\langle 111 \rangle$ NV axis, the sensitivity is
\begin{equation}
\eta = \frac{\sqrt{3}}{\gamma|\partial S/\partial \Omega|}\frac{\sigma_S}{\sqrt{2 f_{\textrm{ENBW}}}}.
\end{equation}
Figure~\ref{fig:RabiMagSignalASD}(a) shows a calibrated amplitude spectral density for a Rabi magnetometry measurement with $\Omega = 2\pi \times 42.8$~kHz and $\omega = 2\pi \times 2.863$~GHz. The rms average sensitivity in the 100~-~3307~Hz band is~3.4~pT/$\sqrt{\textrm{Hz}}$, indicating that this magnetometer operates close to the shot-noise limit of~1.34~pT/$\sqrt{\textrm{Hz}}$ (as calculated in the Supplemental Material~\cite{Alsid2021RabiSupplement}).

As a demonstration of our NV-ensemble Rabi magnetometer, we perform proof-of-principle detection of analog and digital modulation signals, which form the basis for several near- and far-field communication protocols~\cite{Kim2017NearFieldComm,Meyer2018}. Application of a small resonant MW field $\delta B_{\textrm{MW}}$ in addition to the MW driving field $B_{\text{MW}}$ modifies the Rabi frequency, which causes a fluorescence change as illustrated in Fig.~\ref{fig:T2rhoandRFscan}(c). Assuming $B_\text{MW}$ and $\delta B_\text{MW}$ lie along the same axis, the resulting Rabi frequency will fall in the range $\gamma(B_{\textrm{MW}}\pm\delta B_{\textrm{MW}})/\sqrt{3}$~\cite{Alsid2021RabiSupplement}. Any amplitude or phase modulation present on $\delta B_{\textrm{MW}}$ will likewise modulate the resulting NV fluorescence. Figure~\ref{fig:RabiMagSignalASD}(b-c) shows an example of a sine-wave amplitude modulation and square-wave phase modulation. For digital modulation in particular, if fluorescence thresholds are associated with a logical 0 and 1, then basic communication protocols, such as binary phase shift keying, can easily be implemented~\cite{Meyer2018}.  


\section{Discussion and Outlook}


In conclusion, we demonstrate a NV-diamond MW magnetometer achieving 3.4~pT/$\sqrt{\textrm{Hz}}$ sensitivity, near the device's shot-noise limit. 
Additional optimization is expected to yield further performance improvements. For example, longer driven coherence times $T_{2\rho}$ can be achieved by increasing the MW homogeneity over the sensor volume through better matching the dielectric resonator and diamond sample geometries. Furthermore, applying additional control fields can suppress noise in the Rabi drive field~\cite{Stark2017CDD,Farfurnik2017CDD,Wang2020coherence}. Finally, irradiation and annealing of the diamond can increase the number of NV centers interrogated while minimally affecting the decoherence time~\cite{Alsid2019}.\par

This technique can be adapted to different high-frequency regimes. A stronger applied static field $B_0$ can allow extension of this sensing technique to both higher and lower frequencies~\cite{Stepanov2015}; an approximately 50~gauss increase in $B_0$ is expected to shift the resonance frequency by $\pm$150~MHz and allow magnetometry without substantial loss of sensitivity. Further increases in $B_0$ are expected to degrade performance, e.g. due to bias field gradients, precluding high-sensitivity Rabi magnetometry down to the MHz regime. However, the spin-locking protocol~\cite{Loretz2013}, where the MW drive Rabi frequency is matched to the frequency of the target field, can allow detection of fields with frequencies in the hundreds of kHz to few MHz with a modest bias field. Sensing arbitrary frequencies, e.g.~in the 100s of MHz, can be accomplished using a new technique, called quantum frequency mixing~\cite{wang2022sensing}.  This protocol mixes an AC bias field with the desired signal field such that the frequency sum or difference matches the NV energy level splitting, also without requiring large bias fields.\par

The demonstrated~pT/$\sqrt{\textrm{Hz}}$ sensitivity shows that, with adequate engineering and optimization, sensors based on NV ensembles are capable of sensing MW magnetic fields with comparable performance to that demonstrated for static and low-frequency fields. The reported sensitivity is orders of magnitude better than previous diamond-based sensors operating in the MW-frequency regime~\cite{wang2015high,Shao2016,Horsley2018}, and these results could position NV ensembles as a practical platform for high-frequency sensing applications, including microwave circuitry imaging~\cite{Anderson2018Imaging} and near-field antenna characterization~\cite{Jomaa2017RFIDAntenna}. Extensions of Rabi magnetometry, such as spin-locking~\cite{Loretz2013} and quantum frequency mixing~\cite{wang2022sensing}, can be implemented in these devices to further widen the application space by opening up new frequency ranges for high-sensitivity diamond-based magnetometry. 


\section*{Acknowledgements}
The authors thank A. Libson, C. Hart, J. T. Oon, and Y. Liu 
for helpful comments and discussions. We thank L. Pham for strain-gradient measurements of the diamond. 
This work was supported in part by the Q-Diamond Grant No. W911NF13D0001. S. A. was supported by the National Science Foundation (NSF) through the NSF Graduate Research Fellowship Program. 

\bibliography{NVMasterBibliography_202106.bib}

\appendix

\end{document}